\documentclass[twocolumn,superscriptaddress,showpacs,prl,aps,amsmath,amssymb,nofootinbib]{revtex4-1}                                        
\usepackage{natbib}
\usepackage{graphicx,color}
\usepackage{amsmath,amssymb}
\usepackage{verbatim}
\usepackage{float}
\usepackage{wasysym}
\usepackage{amssymb,graphicx}
\usepackage{epsfig}
\usepackage{psfrag}
\usepackage{dsfont}
\usepackage{amsfonts}
\usepackage{mathrsfs}
\usepackage{multirow}
\usepackage{times}
\usepackage{bm}
\usepackage{hyperref}
\usepackage{xspace}
\hypersetup{
  colorlinks=true,        
  linkcolor=blue,         
  citecolor=cyan,         
}
\usepackage{pifont}
\usepackage[normalem]{ulem}   
\usepackage{grffile}
\usepackage{soul}
\usepackage{enumerate}
\usepackage{stackengine}

\newcommand{\GA}{\alpha}
\newcommand{\GB}{\beta}
\newcommand{\GG}{\gamma}

\newcommand{\GR}{\rho}

\newcommand{\GT}{\tau}
\newcommand{\GC}{\psi}

\newcommand{\GP}{\phi}

\newcommand{\pd}{\partial}
\newcommand{\be}{\begin{equation}}
\newcommand{\ee}{\end{equation}}

\newcommand{\TUU}[3]{\tilde{#1}^{#2 #3}} 
\newcommand{\TDD}[3]{\tilde{#1}_{#2 #3}}

\newcommand{\fDu}[1]{\stackon[-0.3ex]{$D^{#1}$}{\kern-1.0ex\scalebox{0.7}{$\circ$}}}

\newcommand{\zD}{{\raise1.0ex\hbox{${}^{\ \circ}$}}\!\!\!\!\!D}

\def\QEQ{{%
    \setbox0\hbox{$I$}%
    \rlap{\hbox to \wd0{\hss--\hss}}\box0
}}

\begin{document}

\title{Magnetohydrodynamic simulations of self-consistent rotating neutron stars with mixed poloidal 
and toroidal magnetic fields}

\author{Antonios Tsokaros}
\affiliation{Department of Physics, University of Illinois at Urbana-Champaign, Urbana, IL 61801}
\email{tsokaros@illinois.edu}

\author{Milton Ruiz}
\affiliation{Department of Physics, University of Illinois at Urbana-Champaign, Urbana, IL 61801}
\author{Stuart L. Shapiro}
\affiliation{Department of Physics, University of Illinois at Urbana-Champaign, Urbana, IL 61801}
\affiliation{Department of Astronomy \& NCSA, University of Illinois at Urbana-Champaign, Urbana, IL 61801}
\author{K\=oji Ury\=u}
\affiliation{Department of Physics, University of the Ryukyus, Senbaru, Nishihara, Okinawa 903-0213, Japan}

\date{\today}

\begin{abstract}
We perform the first magnetohydrodynamic simulations in full general relativity of
self-consistent rotating neutron stars (NSs) with ultrastrong mixed poloidal and toroidal
magnetic fields. The initial uniformly rotating NS models are computed
assuming perfect conductivity, stationarity, and axisymmetry.
Although the specific geometry of the mixed field configuration can delay or
accelerate the development of various instabilities known from analytic
perturbative studies, all our models finally succumb to them. Differential
rotation is developed spontaneously in the cores of our magnetars which, after
sufficient time, is converted back to uniform rotation.  The rapidly rotating
magnetars show a significant amount of ejecta, which can be responsible for
transient kilonova signatures. However no highly collimated, helical magnetic
fields or incipient jets, which are necessary for gamma-ray bursts,  arise at
the poles of these magnetars by the time our simulations are terminated.
\end{abstract}

\maketitle

\textit{Introduction.}\textemdash
Neutron stars are not only the densest objects in the Universe, but in some cases
they possess a magnetic field billions of times larger than the strongest magnet
on Earth. These so called magnetars \cite{Duncan1992,ThompsonDuncan1993,ThompsonDuncan1996}
have magnetic fields that surpass the quantum electrodynamic threshold
of $4\times 10^{13}\,\rm G$ and are responsible for many exotic phenomena,
such as vacuum birefringence, photon splitting, and the distortion of atoms
(see \cite{Harding2006} for a review). They are invoked in order to explain
the large bursts of gamma-rays and X-rays in soft-gamma repeaters \cite{Kouveliotou1998} 
or the related anomalous X-ray pulsars \cite{Gavriil2002}.

Magnetars are also naturally produced after the merger of two NSs via
the instigation of various magnetic instabilities such as the Kelvin-Helmholtz
instability \cite{Price:2006fi,Anderson:2008zp,Kiuchi:2015sga,Kiuchi:2015qua,Dionysopoulou:2015tda}, the
magnetorotational instability (MRI) \cite{Shibata:2005mz,dlsss06a,Siegel:2013nrw,Kiuchi:2015qua}, or
magnetic winding \cite{Baumgarte:1999cq,Shapiro:2000zh,Kiuchi:2015sga}. Even if the two
NSs that compose the binary system have magnetic fields of the order
of $\sim 10^{11}\,\rm G$, when they merge because of the aforementioned
mechanisms the magnetic field reaches magnetar strengths and beyond on a
dynamical timescale. This was first demonstrated with the very high resolution
studies in \cite{Kiuchi:2015qua,Kiuchi:2015sga} where an initial magnetic field
of $10^{13}\,\rm G$ was amplified to  $\gtrsim 10^{15} \,\rm G$, with local
values reaching $\sim 10^{17}\,\rm G$, $5\,\rm ms$ following merger.  Similar
results have been reported more recently in \cite{Aguilera-Miret:2020dhz} where
an even larger amplification was achieved. The existence of this ultrastrong
magnetic field is one of the most crucial factors for the realization of
multimessenger astronomy. According to our current understanding, the merger
of the two NSs in event GW170817 \cite{TheLIGOScientific:2017qsa}
produced such a magnetar that was instrumental for the creation of the sGRB 
\cite{Monitor:2017mdv} (possibly following its delayed collapse), and the kilonova \cite{Valenti:2017ngx,Metzger2018}
that followed.

Despite the large amount of research in analytical and perturbative
magnetohydrodynamics, self-consistent general relativistic solutions of the
Einstein-Maxwell-Euler system are still in their infancy. In \cite{Bocquet1995}
self-consistent equilibria have been obtained with only poloidal magnetic
field, while a different formalism was employed in
\cite{Kiuchi2008,Kiuchi2009a} to obtain self-consistent equilibria with only
toroidal magnetic fields. Other authors have explored such solutions in great
detail \cite{Cardall2001,Yasutake2010,Frieben2012,Chatterjee2014,Franzon2015}
while in \cite{Pili2014,Bucciantini2015,Pili2017} solutions with mixed poloidal and
toroidal magnetic fields were obtained with the price of greatly reducing the
number of Einstein equations solved. On the other hand equilibrium solutions
are not necessarily stable, and indeed, the first general
relativistic MHD simulations with either purely toroidal magnetic fields
\cite{Kiuchi2008b} or purely poloidal magnetic fields
\cite{Ciolfi2011,Lasky2011,Ciolfi2012,Lasky2012} confirmed the unstable nature of these
solutions predicted decades ago
\cite{Tayler_1957,Tayler1973,Wright1973,Markey1973,Markey1974,Flowers1977}.  In
\cite{Ciolfi2011,Lasky2011,Ciolfi2012,Lasky2012} the initial
conditions were based on the self-consistent poloidal solutions of
\cite{Bocquet1995}. In all cases the Cowling approximation was used, i.e. the
Einstein field equations were not evolved but only the MHD equations on the fixed
initial background.  In \cite{Kiuchi2008b} the initial toroidal conditions were
those of \cite{Kiuchi2008} and an axisymmetric GRMHD simulation was employed.

The stability of a pulsar magnetic field, as well as the recent results by
NICER \cite{Miller_2019,Riley:2019}, demand a more sophisticated modelling of a
NS magnetic field. As a first step we go beyond the previous works
above by constructing rotating, magnetized equilibria with mixed ultrastrong
poloidal and toroidal components and evolve them in full GRMHD in order to
assess their evolutionary fate.  Our initial data are constructed with the
magnetized, rotating NS libraries of the \textsc{cocal} code
\cite{Uryu:2014tda,Uryu:2019ckz}, where the \textit{whole} set of
Einstein-Maxwell-Euler system is solved to construct models under the assumptions of perfect
conductivity, stationarity, and axisymmetry. These models are then evolved
using the Illinois GRMHD code \cite{Etienne:2010ui} in full general relativity.
The salient characteristics of our simulations are summarized in the
\textit{Results} section below, while details on the construction of the
self-consistent models, as well as our choices for the simulations, are provided
in the Supplemental Material. Here and throughout we adopt units of 
$G=c=M_\odot=1$, unless otherwise noted.

\setlength{\tabcolsep}{5pt}                                                      
\begin{table*}[ht]
\caption{The magnetar models evolved in this work. Columns are: the model name,
the polytropic index, the central rest-mass density in $g/cm^3$, the gravitational mass,
the period, the ratio of polar to equatorial radii, the rotational kinetic
energy, the total magnetic energy, the toroidal magnetic energy, the poloidal
magnetic energy, the dynamical timescale
($1/\sqrt{\GR_0}$), and the Alfv\'en  timescale. $|\mathcal{W}|$ denotes the
gravitational energy, while the * denotes that this model collapses to a black
hole. }
\label{tab:models}
\begin{tabular}{ccccccccccccc}
\hline\hline
Case & $\Gamma$ & $\rho_{0c}$ & $M$  & $P/M$  & $R_p/R_e$ & $\mathcal{T}/|\mathcal{W}|$ & $\mathcal{M}/|\mathcal{W}|$  & 
$\mathcal{M}_{\rm tor}/|\mathcal{W}|$ & $\mathcal{M}_{\rm pol}/|\mathcal{W}|$ & $t_d/M$ & $t_A/M$   \\
     & & $(\times 10^{15})$ &  &  &   & $(\times 10^{-2})$ & $(\times 10^{-2})$ & $(\times 10^{-4})$ & $(\times 10^{-2})$  & $(\times 10^{18})$  & &   \\
\hline
A1 &  $2$ & $1.072$ & $1.385$ & $173.0$ & $0.7$   & $4.531$ & $3.026$  & $0$      & $2.970$   & $17$ & $56$   \\
A2 &  $2$ & $1.072$ & $1.366$ & $169.3$ & $0.7$   & $4.871$ & $1.632$  & $7.863$  & $1.525$   & $18$ & $70$  \\
A3 &  $2$ & $1.072$ & $1.362$ & $172.0$ & $0.7$   & $4.730$ & $1.794$  & $8.876$  & $1.669$   & $18$ & $61$  \\
A4 &  $2$ & $1.072$ & $1.359$ & $175.3$ & $0.7$   & $4.568$ & $1.983$  & $8.707$  & $1.852$   & $18$ & $47$  \\
A5 &  $2$ & $1.072$ & $1.197$ & $769.3$ & $0.925$ & $0.2612$& $1.709$  & $7.492$  & $1.632$   & $20$ & $45$   \\
A6*&  $2$ & $2.225$ & $1.586$ & $90.77$ & $0.6$   & $5.055$ & $0.1624$ & $0.5361$ & $0.1504$  & $11$ & $126$  \\
A7 &  $3$ & $1.225$ & $1.592$ & $119.1$ & $0.7$   & $4.043$ & $4.399$  & $17.81$  & $4.134$   & $14$ & $18$    \\
\hline\hline
\end{tabular}
\end{table*}

\textit{Initial data.}\textemdash 
We construct the initial magnetized equilibria, models A1-A7 in Table
\ref{tab:models} by solving the Einstein equations, Maxwell's
equations, and ideal MHD equations self-consistently under the
assumptions of stationarity and axisymmetry. 
\begin{figure} 
\begin{center}
\includegraphics[width=0.99\columnwidth]{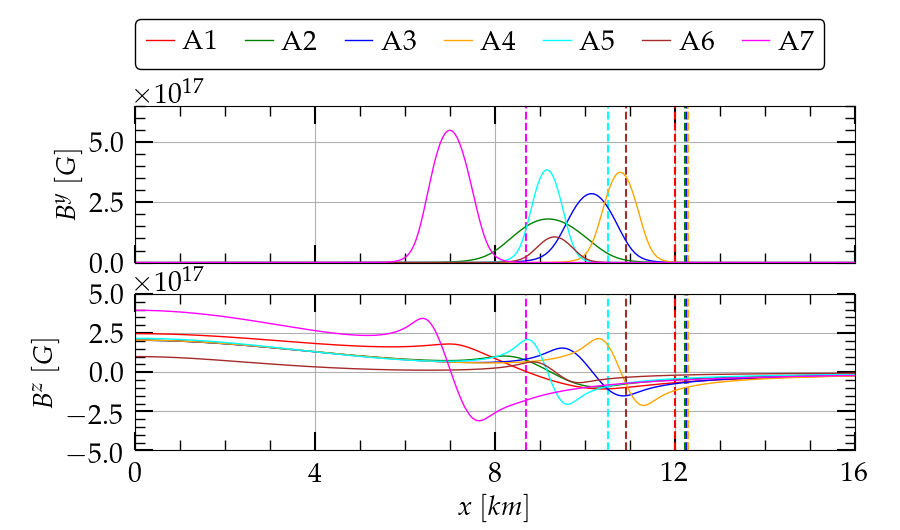}

\vspace{-0.1cm}

\includegraphics[width=0.99\columnwidth]{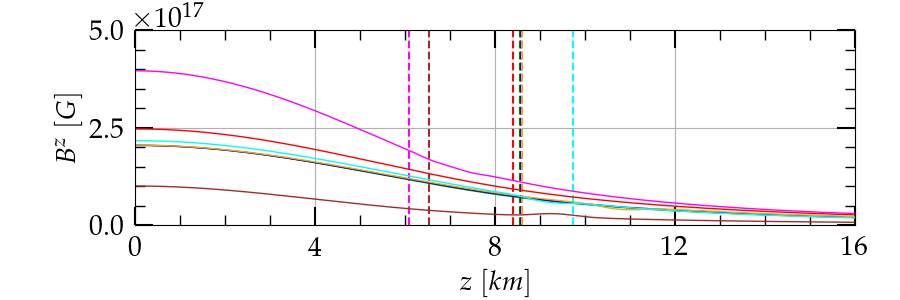}
\caption{Initial magnetic field strength along the x and z axes for all magnetars A1-A7,
where z is the rotational axis.
Vertical dashed lines show the corresponding NS radii. The toroidal magnetic field ($B^y$) is
concentrated in a region below the NS surface. Note that the lines (radii) for A1-A3 closely
overlap.}
\label{fig:B0}
\end{center}
\end{figure}
All of our models use a polytropic equation of state with $\Gamma=2$, except
the last model that has $\Gamma=3$. Model A6 is supramassive \cite{Cook:1993qj},
while all others are normal NSs. Models A1-A4 are rapidly rotating NSs with the
same central density $\rho_{0c}$ and polar to equatorial radius deformation $R_p/R_e$, but with the ratio
of toroidal to poloidal B-field energies $\mathcal{M}_{\rm
tor}/\mathcal{M}_{\rm tor}$ changed systematically.  Model A5 is a slowly
rotating NS whose parameters that determine the B-fields are the same as in
model A4.  Model A6 is close to the mass-shedding limit curve and the maximum
mass of unmagnetized, uniformly rotating equilibrium (to the left).  Finally
model A7 is a moderately rotating normal NS. All magnetars have magnetic energy
which is at most $4.4\%$ of their gravitational potential energy. Most of this
energy is poloidal in nature (expressions of how these energies are computed
are given in \cite{Uryu:2019ckz}) since the toroidal magnetic field is confined
to a region below the surface of the NS, therefore its volume is much
smaller than the corresponding one for the poloidal field, which extends to
infinity. However, the maximum values of the toroidal and the poloidal
magnetic fields are of the same order. This can be seen in Fig. \ref{fig:B0}
where we plot various components of the magnetic field (toroidal is $B^y$)
across the x and z axes. Vertical dashed lines signify the magnetar radii. In
the last two columns of Table \ref{tab:models} we show the dynamical and
Alfv\'en timescales ($t_A = R_e \sqrt{4\pi\rho_0}/B$, where $B$ is the value of the
magnetic field at the NS center) of our models. A more precise estimate of the
Alfv\'en timescale based on the relativistic formula is given in the
Supplemental Material, and broadly agrees with the timescales of  Table
\ref{tab:models}. All models have been evolved for $10-20$ Alfv\'en
times, with the longest being magnetar A7 ($\sim 20 t_A$).

\begin{figure*} 
\begin{center} 
\includegraphics[width=0.5\columnwidth]{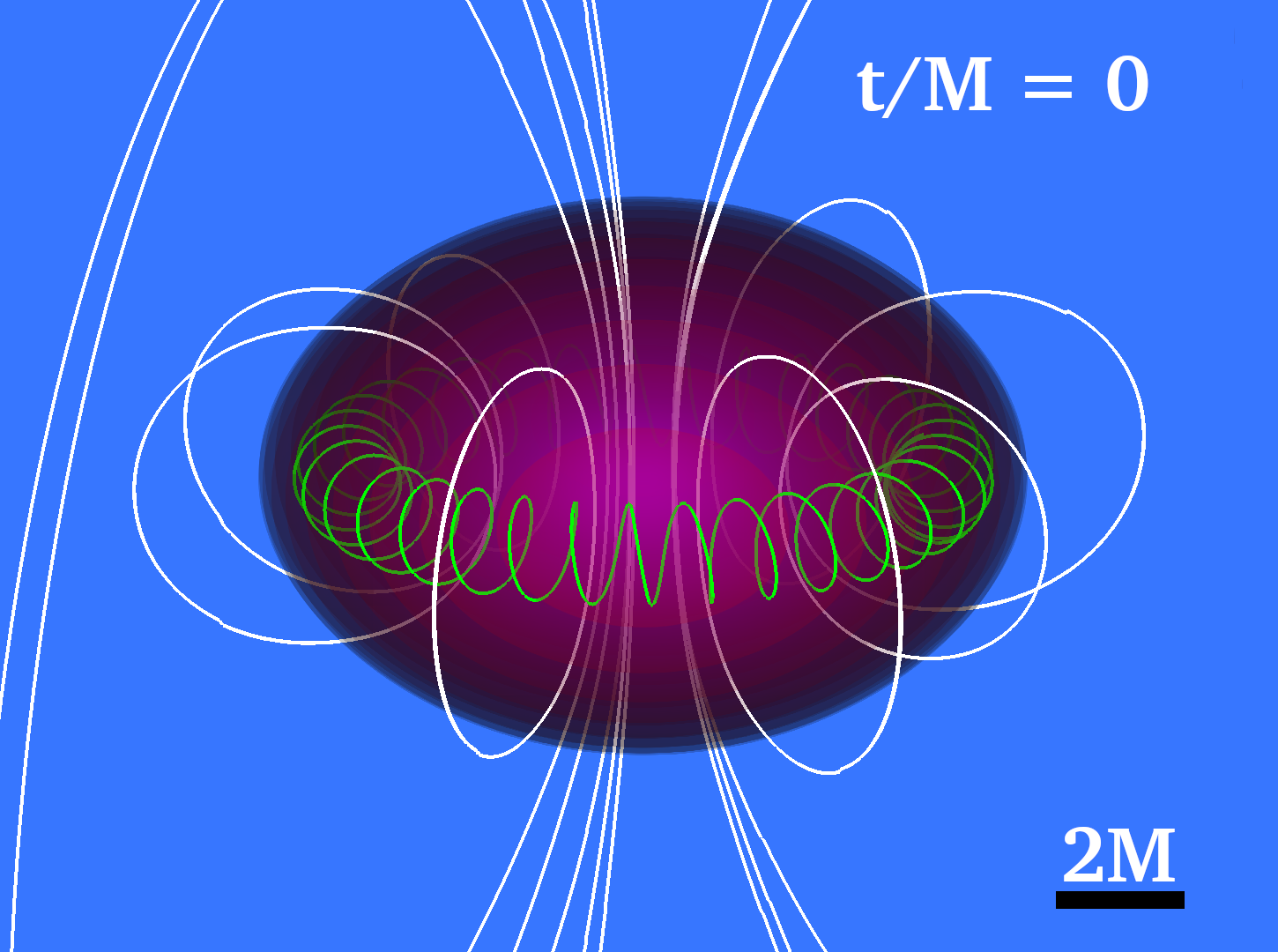}
\includegraphics[width=0.5\columnwidth]{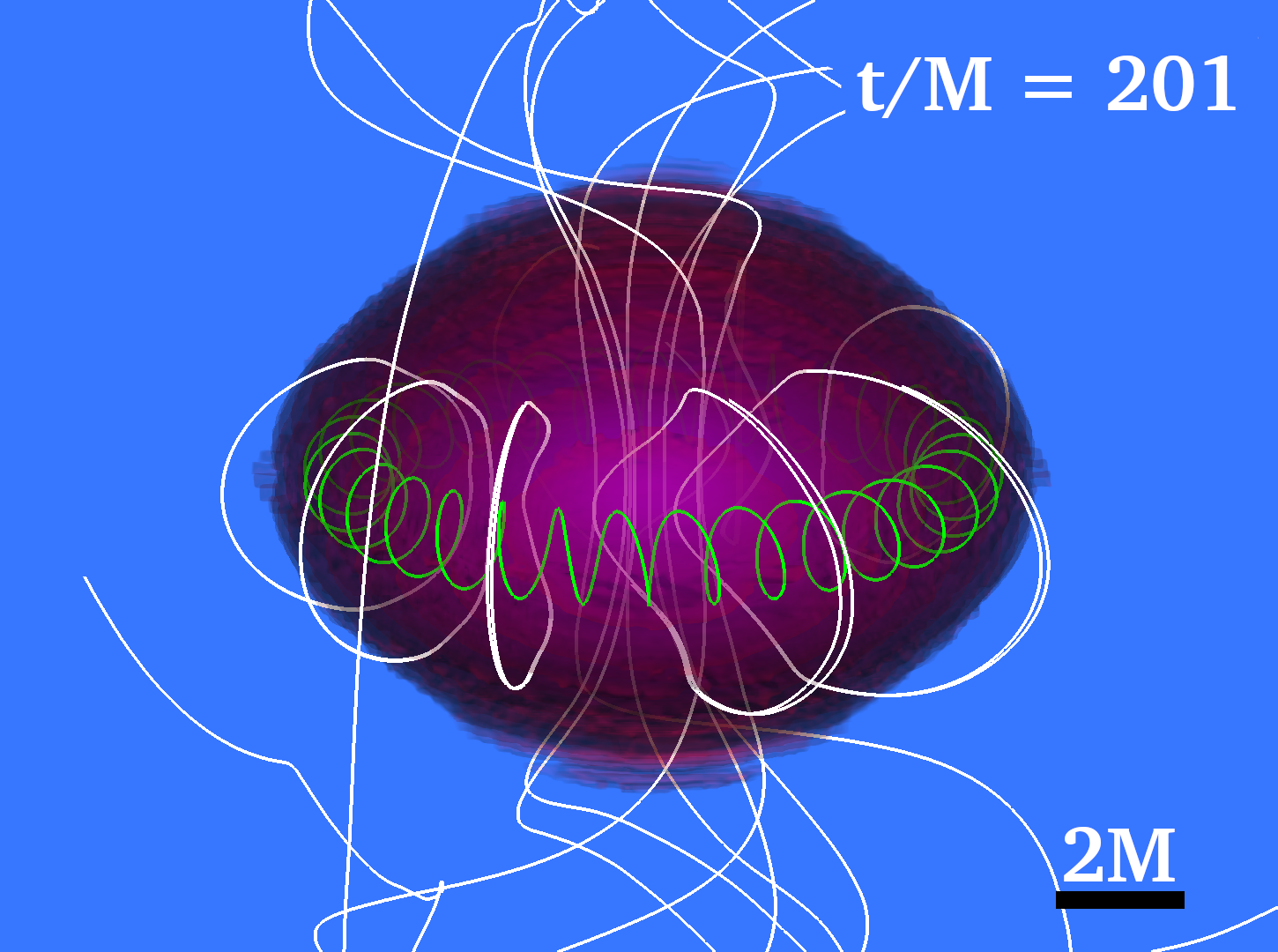}
\includegraphics[width=0.5\columnwidth]{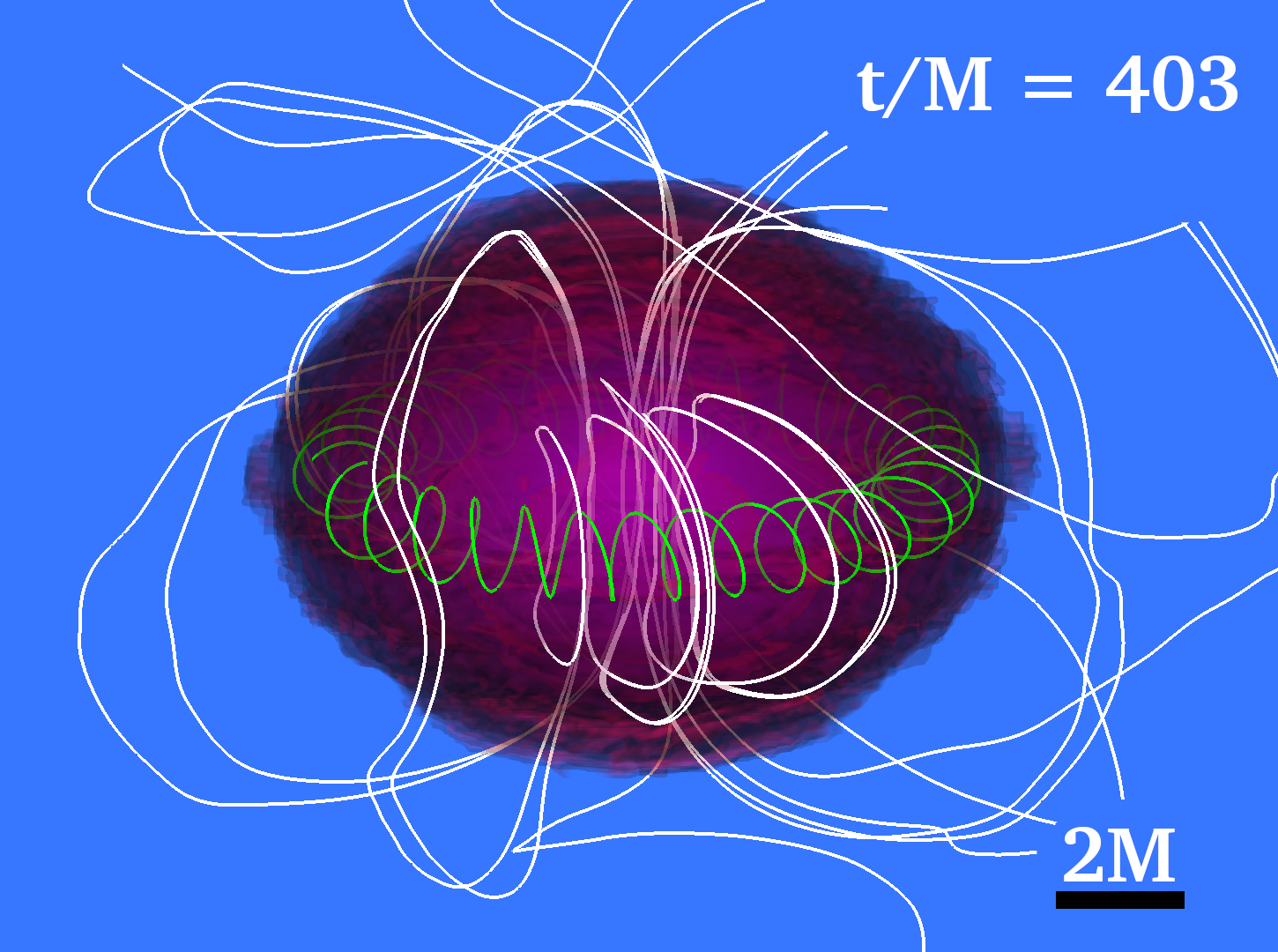}
\includegraphics[width=0.5\columnwidth]{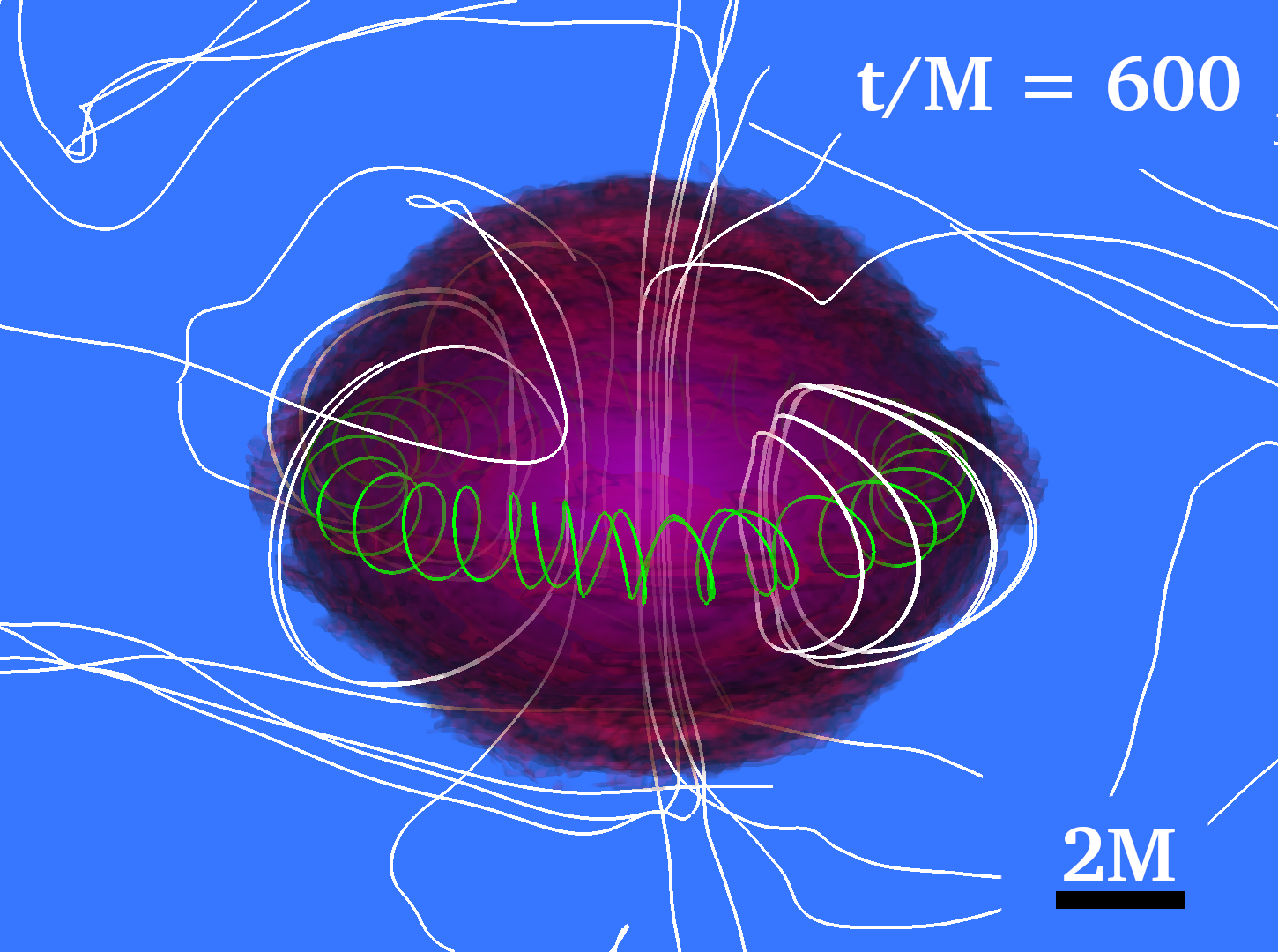}

\vspace{0.1cm}

\includegraphics[width=0.5\columnwidth]{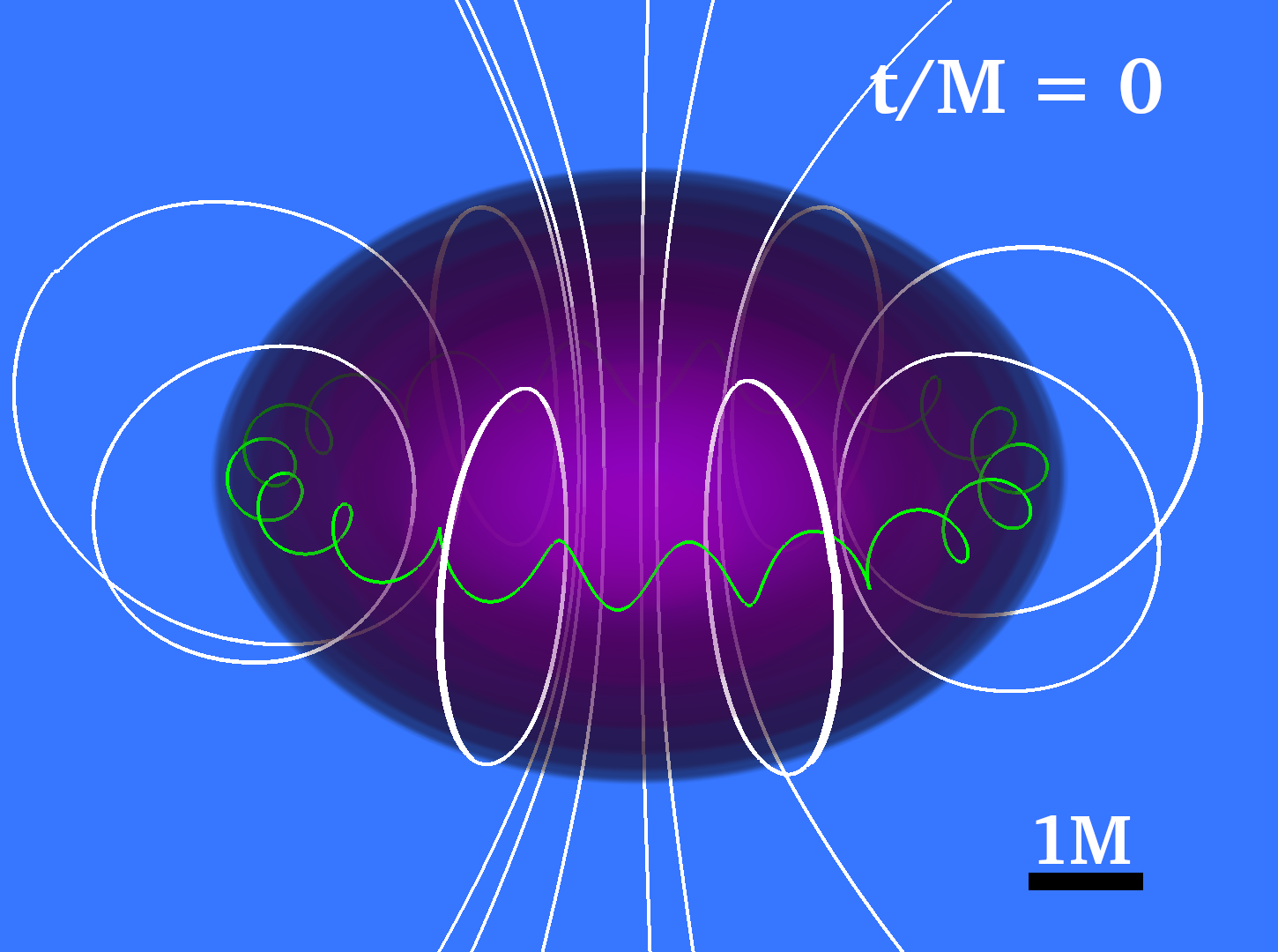}
\includegraphics[width=0.5\columnwidth]{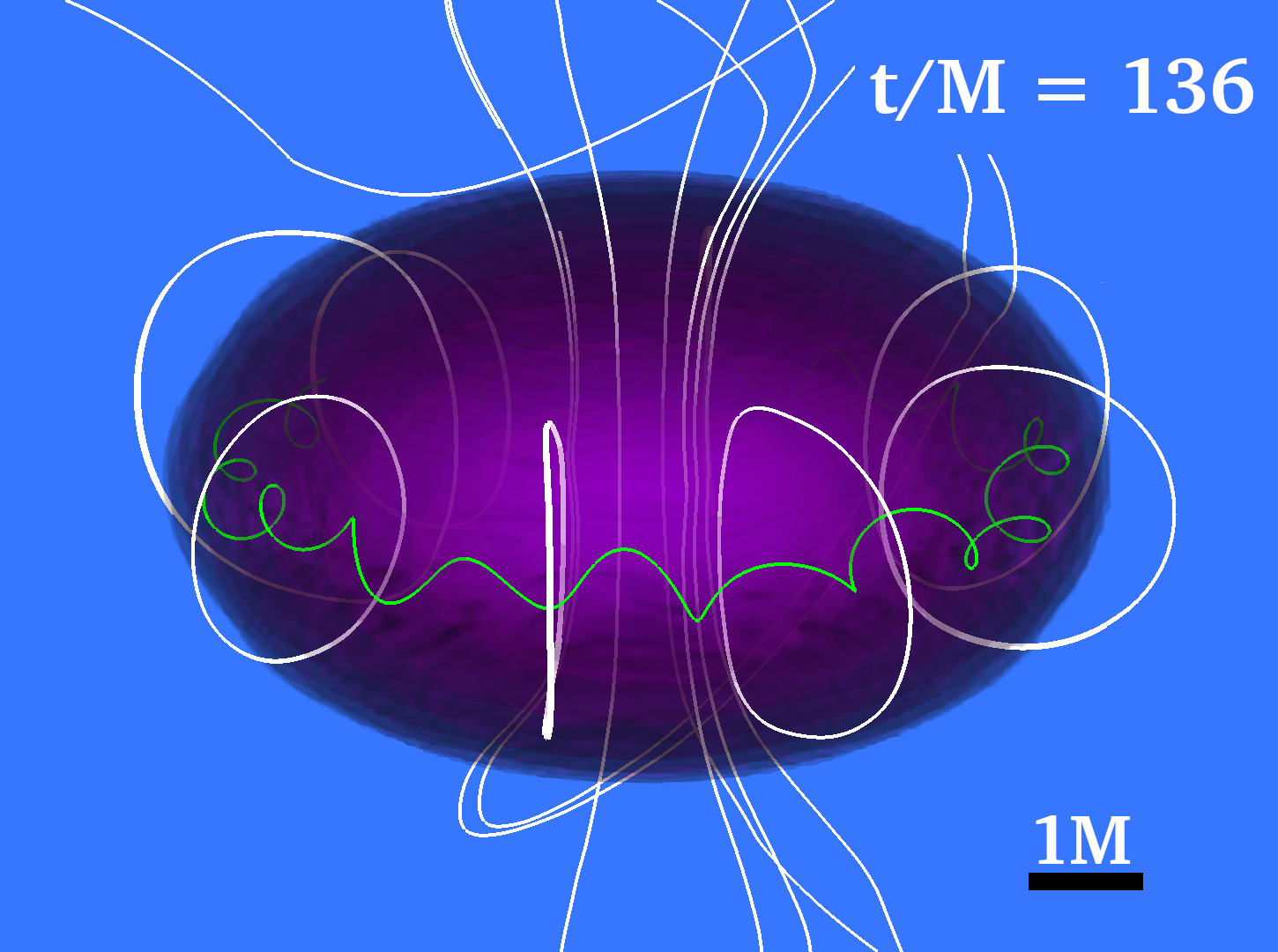}
\includegraphics[width=0.5\columnwidth]{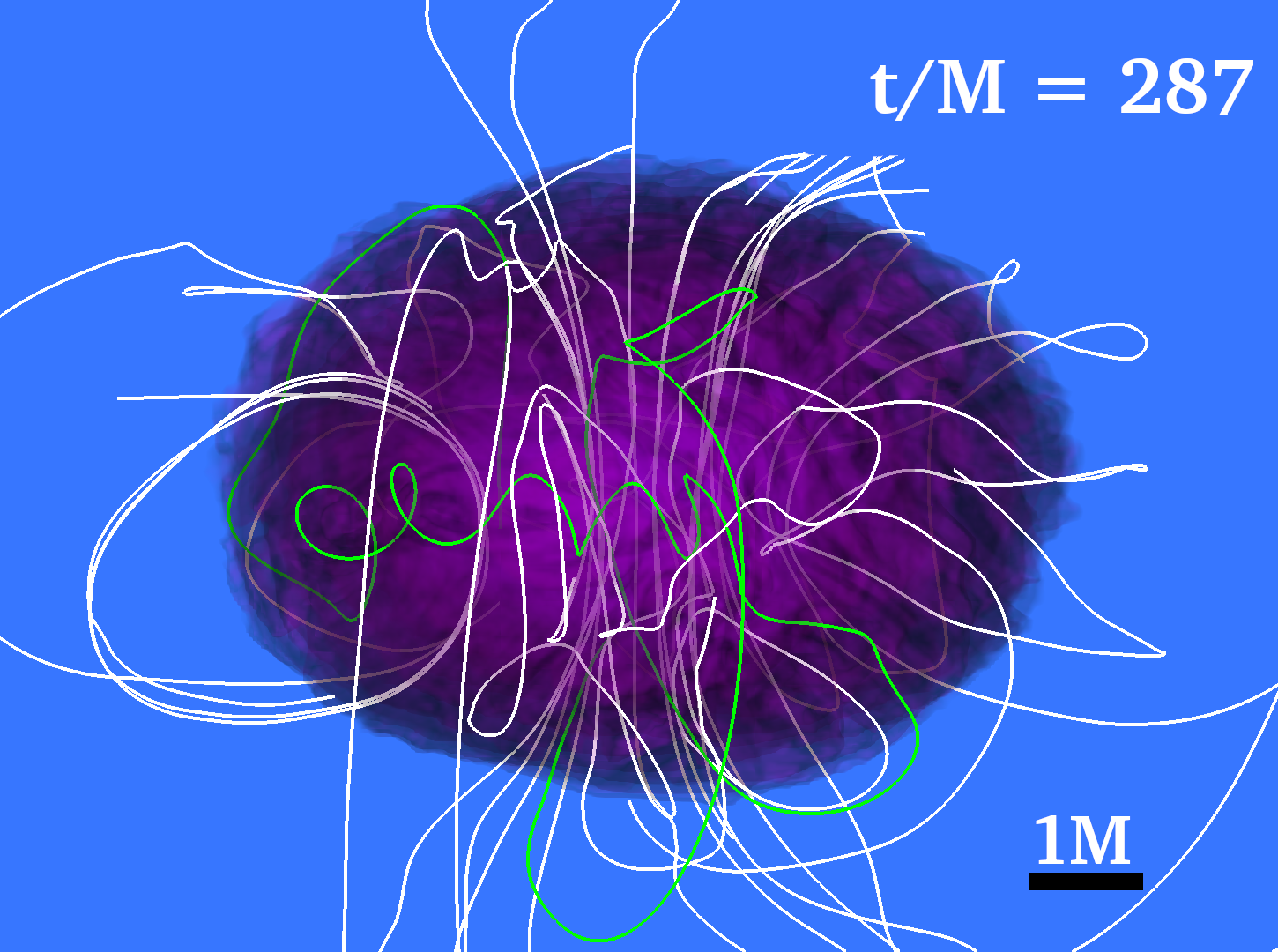}
\includegraphics[width=0.5\columnwidth]{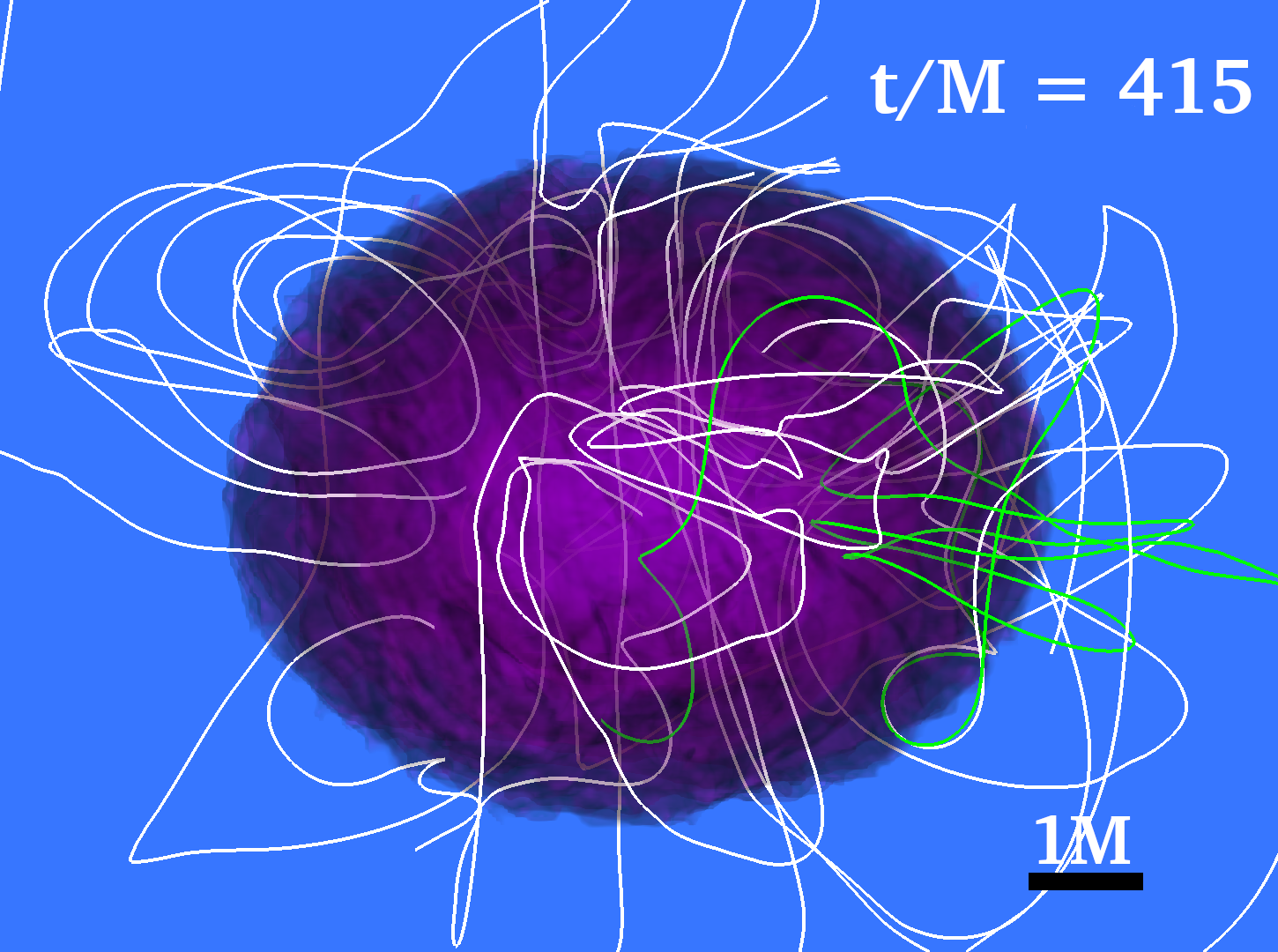}
\caption{Three-dimensional renderings of model A2 (top row) and A7 (bottom row)
at four different instances of time.  White lines show the poloidal field lines
while green lines show a poloidal + toroidal one. The tighter the coil of the
helix, the smaller the toroidal magnetic field.}
\label{fig:A27}
\end{center}
\end{figure*}

\begin{figure*} 
\begin{center}
\includegraphics[width=2.0\columnwidth]{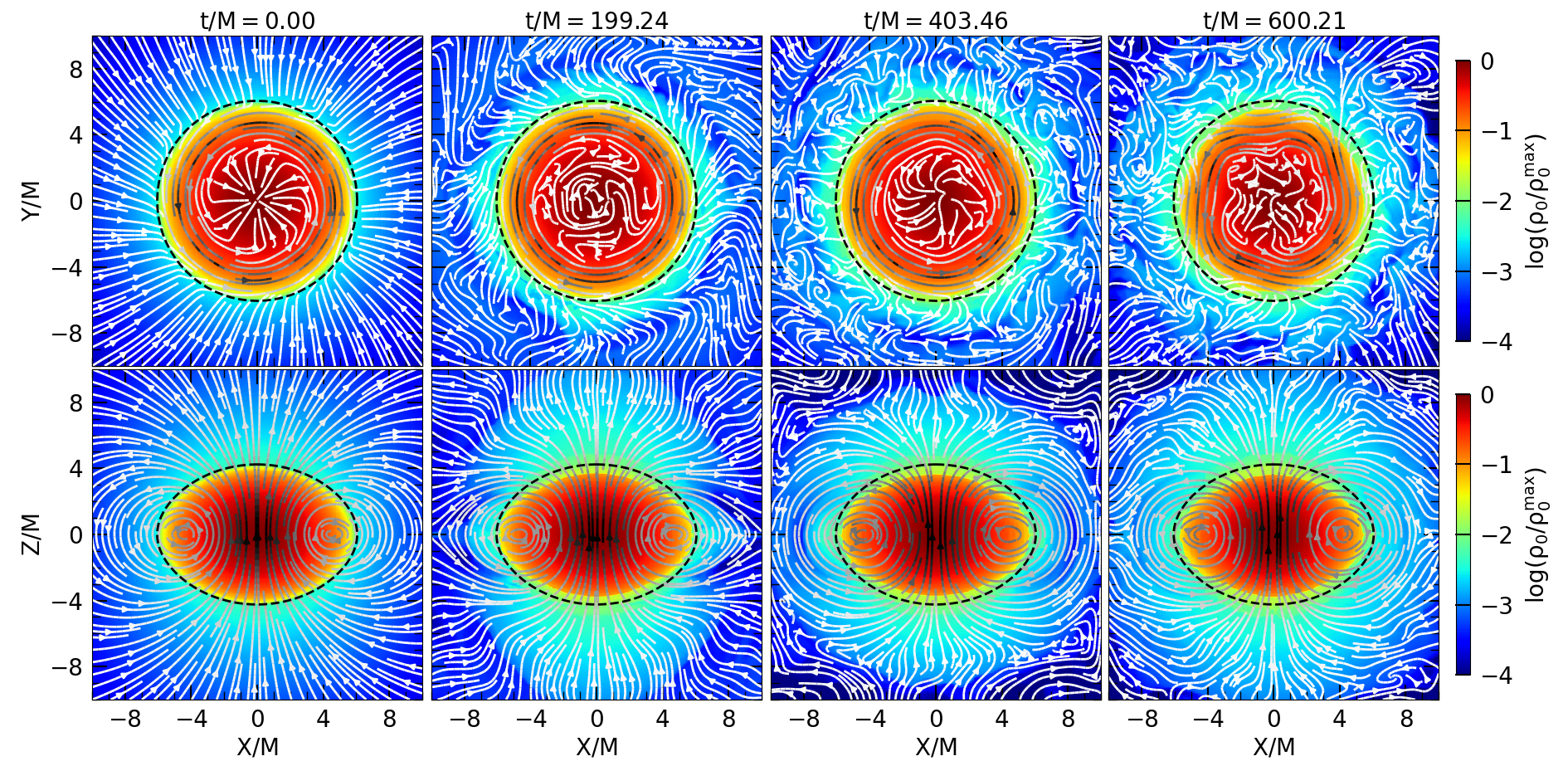}
\caption{Rest-mass density and magnetic field lines for model A2 on the
equatorial and meridional planes at four time instances. Dark field lines
signify a stronger magnetic field.}
\label{fig:A2}
\end{center}
\end{figure*}

\textit{Evolutions.}\textemdash
Magnetars A1-A7 are evolved using the Illinois GRMHD moving-mesh-adaptive-refinement
code (see e.g.~\cite{Etienne:2010ui}) using the settings described in
\cite{Ruiz:2020zaz}, and summarized in the Supplemental Material. In all our
simulations we use high resolution, with the finest refinement level having
$\Delta x_{\rm min}=87$ m for the $\Gamma=2$ models, having radii $10.5-12.3$
km, and $\Delta x_{\rm min}=72$ m for the $\Gamma=3$ model, which has a radius
of $8.7$ km.

\textit{Results.}\textemdash
The behavior of our magnetized neutron stars can be broadly described by the
following characteristics: (i) Large radial density oscillations, especially
for the rapidly rotating magnetars. (ii) Uniform rotation is destroyed in the
core of the stars, which at instances becomes counterrotating. (iii) The NSs
remain axisymmetric to a large degree.  (iv) The toroidal magnetic field is the
first to become unstable.  (v) The timescale of the
instability is longer for smaller toroidal magnetic field strengths, although
the strength is comparable in all cases. Models with the strongest toroidal
B-field exhibit a density dip inside the star (as described in
\cite{Uryu:2019ckz}), are the most unstable and develop a ``gear''-like shape at
the NS surface. (vi) All our models develop the ``varicose'' and ``kink''
instabilities \cite{Markey1973,Tayler1973,Wright1973}. 

In Fig. \ref{fig:A27} we show 3D renderings of model A2 (top row) which has
proven to be the most stable, and model A7 (bottom row) which has the strongest
toroidal magnetic field at 4 different instances.  Also in Fig. \ref{fig:A2}
along with the density profile we show  the poloidal and toroidal field lines
on the meridional and equatorial planes respectively at 4 instances for model
A2. After approximately $\sim 10 t_A$ model A2 preserves broadly both its shape
as well as the geometry of its toroidal and poloidal magnetic fields. In  Fig.
\ref{fig:A27} green lines signify the combined poloidal and toroidal magnetic
field while in Fig. \ref{fig:A2} black field lines signify regions of
strong magnetic field. On the other hand model A7 after  $\sim 10 t_A$ exhibits
turbulent motion on its surface, together with large density oscillations close
to the surface and at $\pm 45^\circ$ from the equatorial plane. By that time
the toroidal geometry of the magnetic field is lost and the kink instability is
fully developed. It has been suggested that the instability can trigger 
giant magnetar flares \cite{ThompsonDuncan1996} and may be accompanied by a
change in the mass quadrupole moment that can potentially lead to detectable
gravitational waves \cite{PhysRevD.83.104014,PhysRevD.83.081302}. In our 
simulations our stars preserve their general (hydrostatic) axisymmetry
and we did not observe any significant nonaxisymmetric mode growth that can
lead to appreciable gravitational wave emission, in accordance with 
\cite{Ciolfi2012,Zink2012,Lasky2012}.

\begin{figure} 
\begin{center}
\includegraphics[width=0.99\columnwidth]{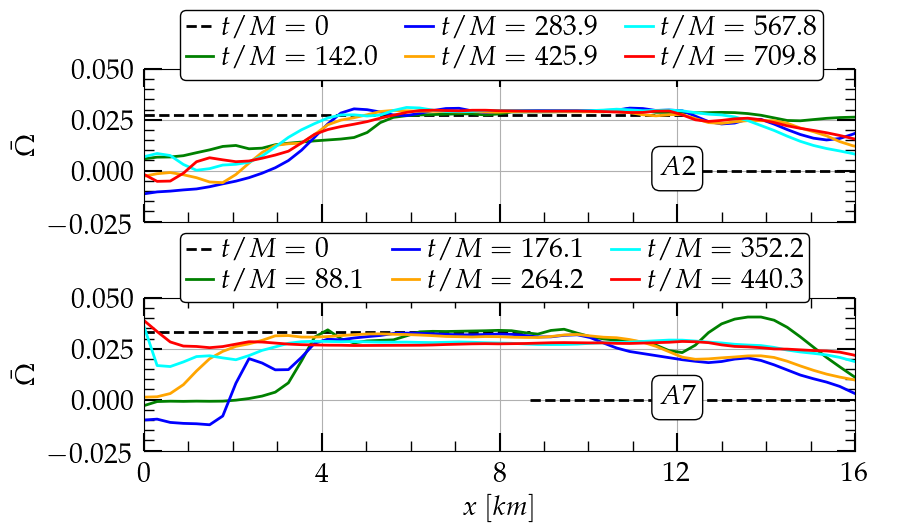}
\caption{Azimuthally averaged angular velocity in the equatorial plane inside the magnetars A2 (top panel) 
and A7 (bottom panel) at 6 instances.}
\label{fig:omega_A27}
\end{center}
\end{figure}

Figure \ref{fig:omega_A27} shows the azimuthally averaged angular velocity
$\bar{\Omega}(r)=1/(2\pi)\int_0^{2\pi}u^\phi/u^t d\phi$ in the equator, plotted
along the x-axis, for models A2 (top panel) and model A7 (bottom panel) at 6
different instances.  Although all our models initially are uniformly rotating,
in a couple of dynamical timescales differential rotation arises in their core
at distances within half their radii. One broad characteristic of this
differential rotation is that it resembles the one found after the merger of
two NSs \cite{Uryu:2017obi,Kastaun:2014fna,Hanauske:2016gia,Ruiz:2021qmm}. In
our simulations this behavior has developed spontaneously and is probably
associated with the strong poloidal magnetic field. If the strength of the
poloidal magnetic field is indeed responsible for the angular velocity drop at
the center of the merger remnant (before uniform rotation is restored) that can
have consequences in its evolution.  A second characteristic is that the
angular velocity in the core at various instances drops to zero and even takes
negatives values, which is reminiscent of the behavior found in analytical
models \cite{Shapiro:2000zh}.  Differential rotation generates toroidal
Alfv\'en waves that convert kinetic energy into magnetic field and thermal
\cite{Cook:2003ku} energy. Eventually we expect that the differential rotation
will be washed out and the star will come back to uniform rotation at the end
due to the effective turbulent viscosity induced by MRI, as can be seen for
model A7 in Fig. \ref{fig:omega_A27}.

\begin{figure} 
\begin{center}
\includegraphics[width=0.99\columnwidth]{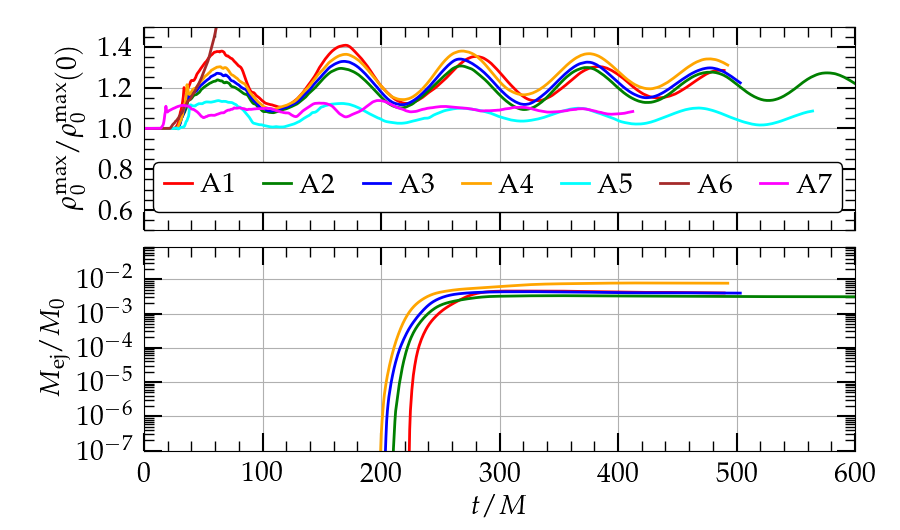}
\caption{Normalized maximum rest-mass density (top panel) and ejecta (bottom panel). Here 
$\rho_0^{\rm max}(0)$ is the initial maximum density and $M_0$ is the initial rest mass.}
\label{fig:rho_ej}
\end{center}
\end{figure}

Models A3, A4, and A5 exhibit a dip in the density profile just below the
NS surface due to the strength of the toroidal magnetic field. This
phenomenon was first found in \cite{Uryu:2019ckz} in mixed poloidal and
toroidal configurations. In our case it is more pronounced in model A4 in
which the pressure becomes zero inside that magnetar, creating a toroidal
electro-vacuum region. These equilibria turn out to be the most unstable since
the radial oscillations in conjuction with the ``varicose'' and ``kink''
instabilities destroy this electro-vacuum on Alfv\'en timescales and create
turbulent-like phenomena on the NS surface. The density profile of
model A2, which was the most stable magnetar, had essentially no such density
dip and therefore no hydrostatic pressure depletion below its surface.

The normalized maximum density of our magnetar solutions is plotted in Fig.
\ref{fig:rho_ej} (top panel) where oscillations of $10-30\%$ are visible. Note
that the largest oscillations are present for the rapidly rotating magnetars
A1-A4 and A6. However, the very slowly rotating model A5 and the moderately
rotating model A7 exhibit oscillations of $\sim 10\%$. The large oscillations
of model A6 are responsible for its collapse to a black hole, since the close
proximity of this magnetar to the maximum supramassive limit \cite{Cook:1993qj}
drives it to the unstable branch \cite{1988ApJ...325..722F} (we have performed
a resolution study to confirm this conclusion).  In the scenario of a binary NS
merger we expect that supramassive remnants close to the maximum mass limit
will be similarly unstable. The timescale of the oscillations in Fig.
\ref{fig:rho_ej} are of the order of an Alfv\'en timescale and are not present
in the absence of the magnetic field.  Indeed, if for the collapsing model A6
we reset the magnetic field to zero, it stays in a quasiequilibrium state and
no collapse is triggered.  In the bottom panel of Fig. \ref{fig:rho_ej} we plot
the ejected material from our simulations. Note that detectable, transient
kilonova signatures powered by radioactive decay of unstable elements formed by
neutron-rich material are expected for ejecta masses $\gtrsim 10^{-3}M_\odot$
\cite{Li:1998bw, Metzger:2016pju}. Using the ejecta of models
$\{A1,A2,A3,A4\}$, and the fitting formulae provided in \cite{Perego:2021dpw},
we infer the peak time $\GT_{\rm peak}$ of the kilonova emission is
$\{5.4,4.6,3.7,5.2\}$ days, the peak luminosity $L_{\rm knova}$ of the ejecta
is $\{1.3,1.2,2.1,2.6\}\times 10^{40}$ ergs/s, and the effective temperature
$T_{\rm peak}$ at the peak is $\{2.4,2.6,2.3,2.1\}\times 10^3$ K, respectively.
Significant ejecta are being created only from our rapidly rotating magnetars
A1-A4, consistent with the suggestion that the magnetic field lines  of a
rotating compact object may accelerate fluid elements due to a
magnetocentrifugal mechanism \cite{1982MNRAS.199..883B}. However no highly
collimated, helical magnetic fields or incipient jets, which are necessary for
gamma-ray bursts,  arise at the poles of these magnetars by the time our
simulations are terminated.

\textit{Discussion.}\textemdash
In this work we presented our latest GRMHD simulations of self-consistent,
ultramagnetized equilibria. We constructed and evolved a diverse set of
magnetars from slowly to rapidly rotating, most having mixed poloidal and toroidal
magnetic fields of comparable strength.  Although the specific geometry of the
mixed field configuration can delay or accelerate the development of various
well-known instabilities \cite{Markey1973,Tayler1973,Wright1973} all our models
finally succumb to them. In addition, our initially uniformly rotating models
develop differential rotation in their cores on a dynamical timescale, similar to
that found in binary NS mergers. To establish exactly how differential rotation
is spontaneously created requires further analysis.  Our models (especially the
rapidly rotating ones) exhibit large radial oscillations in the NS's  F-mode,
and produce significant amounts of ejecta that can power a kilonova. Our
equilibria can be explored further with improved numerical evolution schemes
that will address effects from the artificial atmosphere typically found in all
ideal MHD simulations. In that direction one would employ the equations of
resistive MHD in full GR
\cite{Palenzuela:2008sf,Palenzuela:2012my,Dionysopoulou:2012zv,Dionysopoulou:2015tda} or a scheme
that reliably matches GRMHD to its force-free limit \cite{Paschalidis:2013gma}.
Finally the recent results by NICER \cite{Miller_2019,Riley:2019} are calling
for the development of solutions beyond the large scale dipolar magnetic field
configurations. One important question is how do observed pulsars with their
magnetic fields reach and maintain \textit{uniform} rotation, which is believed
necessary for pulsars to serve as very precise clocks? All of these matters will be the
subject of our future explorations.

\acknowledgements
We thank the Illinois Relativity REU team (H.~Jinghan, M.~Kotak, E.~Yu, and J.~Zhou) for
assistance with some of the visualizations. This work has been supported in part by National Science
Foundation (NSF) Grant PHY-2006066, and NASA Grant 80NSSC17K0070 to the University of
Illinois at Urbana-Champaign, as well as by JSPS Grant-in-Aid for Scientific Research(C) 18K03624 to the 
University of the Ryukyus. This work made use of the Extreme Science and Engineering Discovery
Environment (XSEDE), which is supported by National Science Foundation grant number TG-MCA99S008. This
research is part of the Blue Waters sustained-petascale computing project, which is supported by the
National Science Foundation (awards OCI-0725070 and ACI-1238993) and the State of Illinois. Blue Waters
is a joint effort of the University of Illinois at Urbana-Champaign and its National Center
for Supercomputing Applications. Resources supporting this work were also provided by the NASA High-End
Computing (HEC) Program through the NASA Advanced Supercomputing (NAS) Division at Ames Research Center.

\appendix

\section{Supplemental Material}

\begin{figure} 
\begin{center}
\includegraphics[width=0.99\columnwidth]{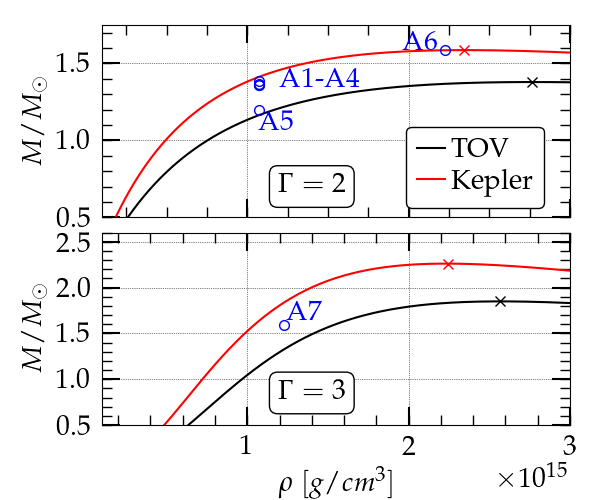}
\caption{Mass versus rest-mass density plots for all our models along with the static and 
uniformly rotating mass-shedding limits. }
\label{fig:Mrho}
\end{center}
\end{figure}

\textit{Initial data.}\textemdash 
Our initial data (blue circles in Fig. \ref{fig:Mrho})
are constructed with the magnetized rotating neutron star
libraries of the \textsc{COCAL} code \cite{Uryu:2014tda,Uryu:2019ckz}.  These
are stationary and axisymmetric equilibrium solutions for the simultaneous
system of Einstein's field equations, Maxwell's equations, and the ideal
magnetohydrodynamic (MHD) equations.

In the 3+1 decomposition, the spacetime metric is written in terms of the lapse
$\GA$, shift $\GB^i$, and the spatial metric $\GG_{ij}$, which is further
decomposed as $\GG_{ij}=\GC^4 \TDD{\GG}{i}{j}$, with
$\TDD{\GG}{i}{j}=f_{ij}+h_{ij}$.  Here $\GC$ is the conformal factor, $f_{ij}$
the flat metric, and $h_{ij}$ the nonflat part of the conformal geometry.  The
introduction of an additional degree of freedom (from 6 components of
$\GG_{ij}$ we now have 7 components for $\GC$ and $h_{ij}$) results to a
constraint for the conformal metric, which we set to be $\tilde{\GG}=1$.
Through combinations of the Einstein equations,  elliptic (Poisson-type)
equations are derived for the 11 components $\{\GA,\GB^i,\GC,h_{ij}\}$.  For
the gauge conditions we use maximal slicing $K=0$, and the Dirac gauge
$\zD_b\TUU{\GG}{a}{b}=0$, where $\zD_a$ is the covariant spatial derivative with
respect to the flat metric $f_{ij}$. A method to impose the latter is described
in \cite{Uryu:2009ye,Uryu:2019ckz} and results in an additional 3 elliptic
equations. Although as shown in \cite{Shibata:2004qz} a waveless condition
$\pd_t \TUU{\GG}{i}{j}=O(r^{-3})$ is sufficient for the metric potentials to
fall off as Coulomb fields, here we use the stronger condition $\pd_t
\TUU{\GG}{i}{j}=0$. Similarly, for the time derivative on the tracefree part of
the extrinsic curvature $\TDD{A}{i}{j}$, as well as on the matter fields, we
use the condition of stationarity. 

As shown in \cite{Gourgoulhon:2011gz}, Maxwell's and the ideal MHD equations
reduce to a single elliptic equation for a master potential called the
relativistic master transfield equation associated with a system of first
integrals on a fluid support.  The transfield equation in the absence of a
meridional flow field becomes the well-known Grad-Shafranov equation
\cite{Shafranov1958,Grad1960,Shafranov1966}.  In our formulation, we do not
reduce the system of Maxwell's and ideal MHD equations to the transfield
equation, but instead, we use the 3+1 decomposition of Maxwell's equations and
solve 4 elliptic equations for the projections of the electromagnetic 1-form
$A_\GA$, subject to the Coulomb gauge $\zD^i A_i=0$, similar to the Dirac gauge
for $\TUU{\GG}{i}{j}$. The ideal MHD condition implies that surfaces of
constant $A_t$ and $A_\GP$ coincide, and therefore these variables are
functions of a single master potential, which is taken to be $A_\GP$ itself.
Then, the first integrals of the MHD-Euler equations and the ideal MHD
conditions become relations to determine the specific enthalpy $h$, the
components of 4-velocity $u^t$ and $u^\phi$, and the components of the current
$j^\alpha$. The latter involves the following arbitrary functions of the
potential $A_\phi$, which are chosen to be
\begin{eqnarray}
\Lambda(A_\GP) & = & -\Lambda_0 \Xi(A_\GP) -\Lambda_1 A_\GP - \mathcal{E},  \label{eq:int1} \\
A_t(A_\GP)  & = & -\Omega_c A_\GP + C_e,  \label{eq:int2} \\
{[\sqrt{-g} \Lambda_\GP]} (A_\GP) & = & \Lambda_{\GP 0} \Xi(A_\GP). \label{eq:int3} 
\end{eqnarray}
In Eqs. (\ref{eq:int1})-(\ref{eq:int3}) 
$\Lambda_0, \Lambda_1$, and $\Lambda_{\GP 0}$ are input parameters that 
control the poloidal and toroidal magnetic field strength, while constants 
$\mathcal{E}$ and $\Omega_c$ are determined during the iteration procedure. 
The former signifies the injection energy \cite{Friedman2012}, while 
the latter is the constant angular velocity of the magnetar. 
Constant $C_e$ controls the net charge of the star, which in our case is zero. $\Xi(A_\GP)$ is 
an integral of the ``sigmoid'' function \cite{Uryu:2019ckz} which is used where 
$A_\GP$ varies on the fluid support, and  
its derivative is written
\be
\Xi'(A_\phi)
= \frac12\left[\,\tanh\left(\frac1{b}
\frac{A_\phi-A_{\phi, \rm S}^{\rm max}}{A_\phi^{\rm max} - A_{\phi, \rm S}^{\rm max}}
-c\right)+1\,\right],  
\label{eq:xi}
\ee
where $A_\phi^{\rm max}$ is the maximum value of $A_\phi$, and $A_{\phi, \rm
S}^{\rm max}$ is the maximum value of $A_\phi$ at the stellar surface.
Parameters $b,c \in [0,1]$ control the width and the position of the sigmoid.
Therefore $\Xi'(A_\phi)$ vary from zero to one in the interval $A_\phi \in
[A_{\phi, \rm S}^{\rm max}, A_\phi^{\rm max}]$.  This guarantees that the
current and the toroidal B field are confined in the star, and the components
of electromagnetic fields extend continuously into the exterior vacuum region.
Together with $\Lambda_0, \Lambda_1$, and $\Lambda_{\GP 0}$ the parameters $b$
and $c$ are reported in Table \ref{tab:param} for the seven models A1-A7
presented here.  

\begin{table}
\caption{Parameters of functions in the integrability conditions Eqs.
(\ref{eq:int1})-(\ref{eq:int3}) and Eq. (\ref{eq:xi}).}
\label{tab:param}
\begin{tabular}{ccccccc}
\hline
Models & $\phantom{-}\Lambda_0$ & $\Lambda_1$ & $\Lambda_{\phi0}$ & $b$ & $c$ & Descriptions \\
\hline
A1 & $\phantom{-}1.6$ & $0.3$ & $0.0$ & $0.2$ & $0.5$ & 
                \multirow{4}{*}{$\left.\begin{array}{l}
                        \\
                        \\
                        \\
                        \\
                \end{array}\right\rbrace$} 
                \multirow{3}{*}{$\begin{array}{l}
                        \\
                        \mbox{Systematic change} \\
                        \mbox{in } B_{\rm tor}^{\rm max}/B_{\rm pol}^{\rm max}
                \end{array}$}   \\                   
A2 & $-0.6$ & $0.3$ & $1.1$ & $0.2$ & $0.5$ &  \\
A3 & $-1.8$ & $0.3$ & $1.7$ & $0.2$ & $0.5$ &  \\
A4 & $-3.0$ & $0.3$ & $2.3$ & $0.2$ & $0.5$ &  \\
A5 & $-3.0$ & $0.3$ & $2.3$ & $0.2$ & $0.5$ & Slow rotation of A4 \\
A6 & $-1.7$ & $0.1$ & $1.7$ & $0.2$ & $0.5$ & Supramassive model \\
A7 & $-0.2$ & $0.3$ & $1.0$ & $0.2$ & $0.5$ & $\Gamma=3$ model \\
\hline
\end{tabular}
\end{table}



\textit{Evolutions.}\textemdash
We evolve the initial data above  using the Illinois GRMHD
moving-mesh-adaptive-refinement code (see e.g.~\cite{Etienne:2010ui}), which
employs the Baumgarte-Shapiro-Shibata-Nakamura formulation of the Einstein’s
equations~\cite{shibnak95,BS} with  puncture gauge conditions~(see~Eq.~(2)-(4)
in~\cite{Etienne:2007jg}). The MHD equations are solved in conservation-law
form adopting high-resolution shock-capturing methods. Imposition of
$\nabla\cdot\vec{B}=0$ during  evolution is achieved by integrating the
magnetic induction equation using a vector potential  $\mathcal{A}^\mu$
\cite{Etienne:2010ui}). The generalized Lorenz gauge \cite{Farris:2012ux} is
employed to avoid the appearance of spurious magnetic
fields~\cite{Etienne:2011re}. We employ a $\Gamma$-law equation of state (EOS),
$P=(\Gamma-1)\,\rho_0\,\epsilon$, with $\epsilon$ the specific  internal
energy, and  allow for shock heating. In models A1-A6 we set $\Gamma=2$, and
$\Gamma=3$ in A7.  To capture one of the properties of the force-free
conditions ($B^2/(8\pi\rho_0)>>1$) that likely characterize the neutron star
exterior, we set a variable and low-density magnetosphere outside the star such
that the magnetic-to-gas pressure ratio is $\beta=P_{\rm mag}/P_{\rm gas}=100$
everywhere \cite{prs15}.  This one-time reset of the low-density magnetosphere
increases the total rest-mass on the entire grid by less than $1\%$, consistent
with the values reported previously (see e.g.
\cite{Ruiz:2016rai,Ruiz:2018wah}).  The ideal GRMHD equations are then
integrated everywhere, imposing on top of the magnetosphere a density floor in
the low density regions similar to
\cite{Ruiz:2020zaz,Ruiz:2016rai,Ruiz:2018wah}.

\begin{figure} 
\begin{center}
\includegraphics[width=0.99\columnwidth]{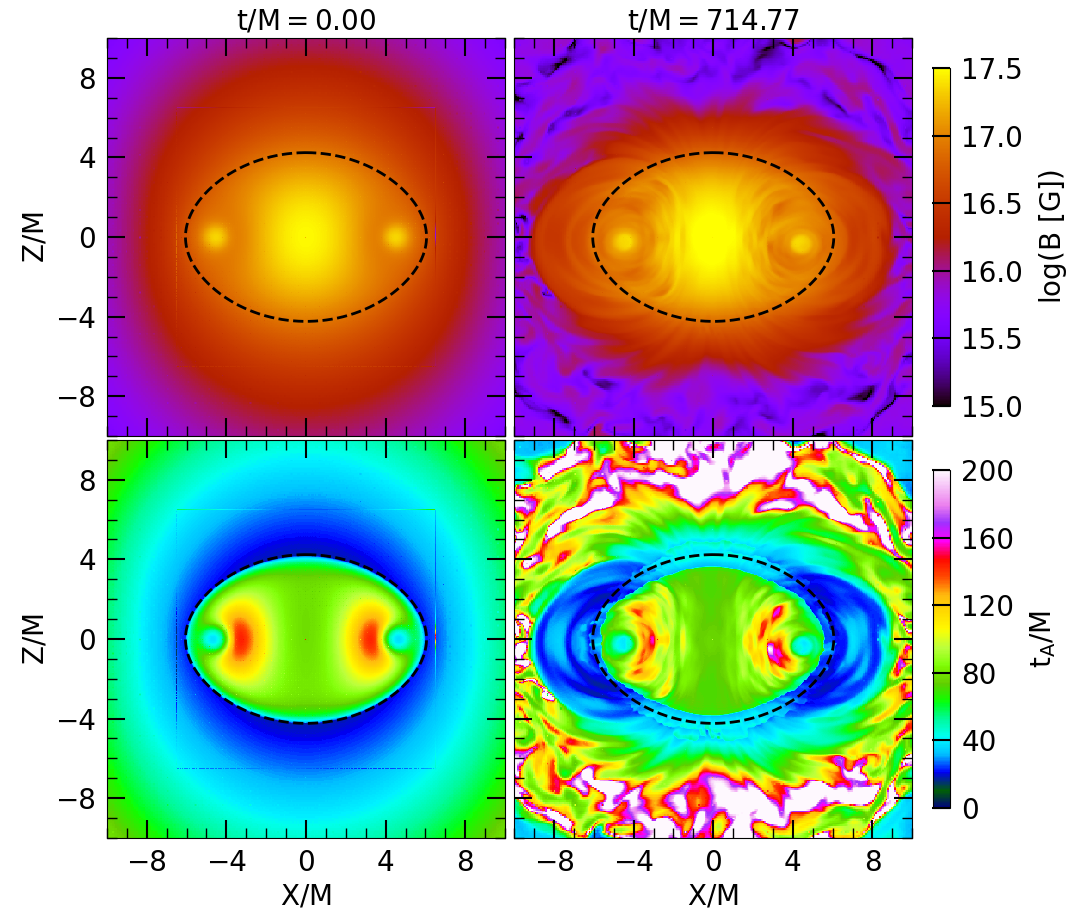}
\caption{Magnetic field strength and Alfv\'en timescale for model A2 at two different instances.}
\label{fig:A2_mag}
\end{center}
\end{figure}
\textit{Magnetic field strength.}\textemdash 
In Fig. \ref{fig:A2_mag} (top row) we show the magnetic field strength for the
model A2 at the initial and final (end of our simulations) moments. Although
the broad structure of the poloidal and toroidal magnetic field is preserved,
it is apparent from the figure that the centers of the toroidal configuration
have moved slightly relative to the equatorial axis as a result of the kink
instability. In the bottom panels we show the Alfv\'en timescale $t_A$ on the
meridional plane based on the relativistic formula $v_A=\sqrt{b^2/(b^2+\rho_0
h)}$, where $v_A$ is the Alfv\'en velocity, $b^\mu=B^\mu/\sqrt{4\pi}$, and
$t_A=R_e/v_A$.  At the end of our simulation the profile of the Alfv\'en
timescale follows closely the initial one.

\bibliographystyle{apsrev4-1}
\bibliography{references}

\end{document}